\documentclass[preprint]{elsarticle}

\usepackage{lineno}
\usepackage[pdfpagemode={UseOutlines},bookmarks=true,bookmarksopen=true, bookmarksopenlevel=0,bookmarksnumbered=true,hypertexnames=false, colorlinks,linkcolor={red},citecolor={blue},urlcolor={red}, pdfstartview={FitV},unicode,breaklinks=true]{hyperref}
\usepackage{adjustbox}
\usepackage{tabularx}
\usepackage{changepage}
\usepackage{mathtools}
\usepackage{amsmath}
\usepackage{graphicx}
\usepackage{txfonts}

\journal{Elsevier}

\begin{document}

\begin{frontmatter}

\title{Revisiting the TeV detection prospects for radio galaxies}


\author[mymainaddress,mysecondaryaddress]{R.\,Angioni\corref{mycorrespondingauthor}}

\cortext[mycorrespondingauthor]{Corresponding author}
\ead{roberto.angioni@ssdc.asi.it}

\address[mymainaddress]{Max-Planck-Institut f\"ur Radioastronomie, Auf dem H\"ugel 69, 53121 Bonn, Germany}
\address[mysecondaryaddress]{Institut f\"ur Theoretische Physik und Astrophysik, Universit\"at
   W\"urzburg, Emil-Fischer-Str. 31, 97074 W\"urzburg, Germany}

\begin{abstract}
Radio galaxies host relativistic jets oriented away from our line of sight, making them challenging targets for Very High Energy (VHE, E$>$100\,GeV) $\gamma$-ray detectors. Indeed, out of $\sim100$ extragalactic sources detected at $E>100$ GeV, only six are radio galaxies, while the great majority are blazars hosting aligned jets. The new Cherenkov Telescope Array (CTA) will provide order-of-magnitude improvements in sensitivity and spectral resolution with respect to the present generation of ground-based $\gamma$-ray observatories, opening new frontiers for high-energy studies of radio galaxies.
In a previous paper, we studied the detection prospects of misaligned jets with the CTA for a sample of sources from the Third \textit{Fermi}-LAT catalog (3FGL). In this work, we complement this study taking advantage of the expanded sample from the Fourth \textit{Fermi}-LAT catalog (4FGL), which includes roughly double the number of sources.
We simulate CTA observations of 41 $\gamma$-ray radio galaxies, extrapolating their \textit{Fermi}-LAT spectrum into the TeV energy range assuming different spectral shapes.
We predict that the CTA will detect eleven new TeV radio galaxies with an observational campaign of 50 hours per source, under the realistic assumption of a spectral cutoff at 0.5\,TeV. This would increase the sample of VHE radio galaxies by a factor of three, and result in the first detection of FR\,II radio galaxies at these energies. By simulating CTA observations with 5 hours exposure, we predict that two \textit{Fermi}-LAT radio galaxies should already be well within reach of current TeV observatories. Finally we have investigated the prospects for a CTA detection of extended VHE emission from the lobe-dominated FR\,I Fornax\,A, and predict that such a detection will be possible for integration times $\gtrsim\!50$ hours.
We conclude that, in line with our previous findings, the CTA will significantly impact our understanding of misaligned jets at TeV energies, allowing us to perform population studies, as well as a comparison between the two main radio galaxy subclasses for the first time in this energy band.
\end{abstract}

\begin{keyword}
\texttt{Galaxies:  active; Galaxies:  nuclei; Galaxies:  jets;Gamma rays:  galaxies}
\end{keyword}

\end{frontmatter}


\section{Introduction}

Active Galactic Nuclei (AGN) hosting relativistic jets are the dominant source class in the extragalactic $\gamma$-ray sky, as revealed by the Large Area Telescope (LAT) on board the \textit{Fermi Gamma-ray Space Telescope} \citep{Atwood2009,4FGL} in the GeV range, and by ground-based Imaging Atmospheric Cherenkov Telescopes (IACTs) in the TeV range~\footnote{See \href{http://tevcat2.uchicago.edu/}{http://tevcat2.uchicago.edu/}}. The vast majority of these sources belong to the subclass of blazars, i.e. they have jets aligned with our line of sight, leading to strong relativistic Doppler boosting and beaming of their radiation. Blazars are typically divided into two subclasses: Flat-Spectrum Radio Quasars (FSRQs), showing broad lines in their optical spectra, and BL Lacs, showing featureless optical spectra. According to the orientation-based unified model~\citep{urry}, the parent population of blazars consists of classic radio galaxies, which can be further classified based on their total radio power and morphology~\citep{fan}. High-power sources ($L_\mathrm{178\,MHz}>2\times10^{24}$ W/Hz/sr), called Fanaroff-Riley type II (or FR\,II), often show a bright core and a one sided kpc-scale jet, terminating in bright compact hot-spots. Low-power sources ($L_\mathrm{178\,MHz}<2\times10^{24}$ W/Hz/sr), i.e. Fanaroff-Riley type I (or FR\,I), show a two-sided jet and edge darkened diffuse lobes. These two classes are believed to be the misaligned counterparts of FSRQs and BL Lacs, respectively. Radio galaxies, together with intermediate objects such as Steep Spectrum Radio Quasars (SSRQs) are collectively refereed to as Misaligned AGN (MAGN), to distinguish them from aligned-jet blazar sources. 

Since radio galaxy jets are not closely aligned with the observer, Doppler boosting plays a much less significant role. Therefore, these sources are typically much more difficult to detect in the $\gamma$-ray band, making up about $\sim1-2\%$ of all extragalactic sources, both in the GeV and in the TeV band. Despite their small numbers, high-energy emitting radio galaxies are a fundamental tool in our understanding of jets, since they provide us with a complementary view with respect to blazars and allow us to test the orientation-based unified models. Moreover, since the emission is not completely dominated by the Doppler-boosted innermost jet, we can observe different emission components. In fact the only extragalactic sources where extended $\gamma$-ray emission has been confidently detected are radio galaxies~\citep{2010Sci...328..725A,2016ApJ...826....1A}.

Currently only six radio galaxies have been detected by IACTs at Very High Energies (VHE, E$>$100 GeV). Detecting additional sources hosting misaligned jets at the highest energies is crucial in order to test our understanding of this phenomenon from different perspectives. The upcoming Cherenkov Telescope Array \citep[CTA,][]{2017arXiv170907997C} will significantly expand our understanding of extragalactic jets at the highest energies, thanks to improvements by an order-of-magnitude in sensitivity and by a factor 2--5 in spatial and spectral resolution. In a previous publication \citep{2017APh....92...42A} we have explored the CTA detection prospects for radio galaxies listed in the Third \textit{Fermi}-LAT source catalog~\citep[3FGL,][]{3fgl}. By simulating CTA observations assuming different extrapolations of the LAT spectra, we found that the CTA is expected to significantly increase the number of TeV radio galaxies. In this work, we expand our previous study by considering the detection prospects for radio galaxies which are newly detected in the Fourth \textit{Fermi}-LAT source catalog~\citep[4FGL,][]{4FGL}.

This paper is organized as follows. In Section~\ref{sec:sample} we define the sample of TeV radio galaxies candidates. In Section~\ref{sec:sim} we outline the procedure used to simulate CTA observations and evaluate the detection significance. In Section~\ref{sec:res} we report and discuss the results of the simulations, the census of the most promising candidates, as well as the driving parameters separating the best candidates from the rest. We also discuss the prospects of detecting extended emission from the lobe-dominated FR\,I radio galaxy Fornax\,A with the CTA. Finally, we summarize our findings and draw our conclusions in Section~\ref{sec:conc}.

\begin{table}[htbp]
\footnotesize
\tabcolsep=1mm
\caption{\label{tab:sample}$\gamma$-ray detected radio galaxies in the 4LAC~\citep{4lac}. Coordinates, redshift and radio flux have been collected from the \href{http://ned.ipac.caltech.edu/}{NASA/IPAC Extragalactic Database (NED)}, while the $\gamma$-ray properties are from the 4FGL~\citep{4FGL}.}
\begin{adjustwidth}{-3.5cm}{-3.5cm}
\begin{center}
\begin{tabular}{llcccccccc}
\hline
\hline
4FGL name & Common name & RA(J2000) & Dec(J2000) & $z$ & FR class & $S_\mathrm{1.4\,GHz}$ & $L_\mathrm{1.4\,GHz}$ & Photon flux & Energy flux\\
\hline
 &  & & & & & Jy & $10^{24}$ W/Hz & $10^{-8}$ ph\,cm$^{-2}$\,s$^{-1}$ & $10^{-6}$ MeV\,cm$^{-2}$\,s$^{-1}$\\
\hline
4FGL\,J0009.7$-$3217 & IC\,1531 & 2.4 & $-$32.28 & 0.026 & I & 0.56 & 0.839 & 0.29 & 1.31\\
4FGL\,J0057.7+3023 & NGC\,315 & 14.45 & 30.35 & 0.017 & I & 0.59 & 0.36 & 0.69 & 2.44\\
4FGL\,J0153.4+7114 & TXS\,0149+710 & 28.36 & 71.25 & 0.022 & I & 0.29 & 0.322 & 0.27 & 2.41\\
4FGL\,J0237.7+0206 & PKS\,0235+017 & 39.42 & 1.98 & 0.022 & I & 0.08 & 0.086 & 0.2 & 0.96\\
4FGL\,J0308.4+0407 & 3C\,78 & 47.11 & 4.11 & 0.029 & I & 7.0 & 13.504 & 0.7 & 4.85\\
4FGL\,J0312.9+4119 & NRAO\,128 & 48.26 & 41.63 & 0.136 & II & 0.67 & 32.936 & 0.85 & 2.57\\
4FGL\,J0316.8+4120 & IC\,310 & 49.18 & 41.32 & 0.019 & I & 0.17 & 0.138 & 0.14 & 1.77\\
4FGL\,J0319.8+4130 & NGC\,1275 & 49.95 & 41.51 & 0.018 & I & 22.8 & 15.926 & 32.4 & 173.0\\
4FGL\,J0322.6$-$3712e & Fornax\,A & 50.67 & $-$37.21 & 0.006 & I & 0.06 & 0.005 & 0.63 & 3.83\\
4FGL\,J0334.3+3920 & 4C\,+39.12 & 53.58 & 39.36 & 0.021 & I & 0.11 & 0.108 & 0.18 & 1.62\\
4FGL\,J0418.2+3807 & 3C\,111 & 64.59 & 38.03 & 0.049 & II & 7.73 & 44.022 & 4.16 & 9.96\\
4FGL\,J0433.0+0522 & 3C\,120 & 68.3 & 5.35 & 0.033 & I & 0.27 & 0.689 & 3.08 & 8.11\\
4FGL\,J0519.6$-$4544 & Pictor\,A & 79.96 & $-$45.78 & 0.035 & II & 6.32 & 17.978 & 0.96 & 2.91\\
4FGL\,J0627.0$-$3529 & PKS\,0625$-$35 & 96.78 & $-$35.49 & 0.055 & I & 1.06 & 7.522 & 1.13 & 10.1\\
4FGL\,J0708.9+4839 & NGC\,2329 & 107.28 & 48.62 & 0.019 & I & 0.68 & 0.57 & 0.1 & 0.79\\
4FGL\,J0758.7+3746 & 3C\,189 & 119.62 & 37.79 & 0.043 & II & 2.6 & 11.168 & 0.13 & 0.89\\
4FGL\,J0931.9+6737 & NGC\,2892 & 143.22 & 67.62 & 0.023 & I & 0.48 & 0.571 & 0.81 & 3.47\\
4FGL\,J1012.7+4228 & B3\,1009+427 & 153.19 & 42.48 & 0.365 & II & 0.08 & 36.19 & 0.17 & 2.34\\
4FGL\,J1116.6+2915 & B2\,1113+29 & 169.16 & 29.26 & 0.047 & I & 1.93 & 9.997 & 0.01 & 0.43\\
4FGL\,J1144.9+1937 & 3C\,264 & 176.27 & 19.61 & 0.022 & I & 5.94 & 6.526 & 0.25 & 2.03\\
4FGL\,J1149.0+5924 & NGC\,3894 & 177.21 & 59.42 & 0.011 & I & 3.51 & 0.905 & 0.25 & 1.5\\
4FGL\,J1230.8+1223 & M\,87 & 187.71 & 12.39 & 0.004 & I & 0.48 & 0.02 & 1.72 & 10.4\\
4FGL\,J1306.3+1113 & TXS\,1303+114 & 196.58 & 11.23 & 0.086 & I & 0.41 & 7.544 & 0.13 & 1.06\\
4FGL\,J1306.7$-$2148 & PKS\,1304$-$215 & 196.68 & $-$21.8 & 0.126 & II & 0.31 & 12.997 & 0.68 & 3.52\\
4FGL\,J1325.5$-$4300 & Centaurus\,A core & 201.37 & $-$43.02 & 0.002 & I & 50.7 & 0.372 & 15.9 & 39.9\\
4FGL\,J1346.3$-$6026 & Centaurus\,B & 206.7 & $-$60.41 & 0.013 & I & 7.9 & 2.951 & 5.83 & 19.0\\
4FGL\,J1443.1+5201 & 3C\,303 & 220.76 & 52.03 & 0.141 & II & 2.54 & 135.456 & 0.13 & 0.93\\
4FGL\,J1449.5+2746 & B2\,1447+27 & 222.37 & 27.78 & 0.031 & I & 0.06 & 0.131 & 0.02 & 0.57\\
4FGL\,J1516.5+0015 & PKS\,1514+00 & 229.17 & 0.25 & 0.052 & I & 0.79 & 5.054 & 1.06 & 2.8\\
4FGL\,J1518.6+0614 & TXS\,1516+064 & 229.69 & 6.23 & 0.102 & II & 0.5 & 13.267 & 0.1 & 1.01\\
4FGL\,J1521.1+0421 & PKS\,B1518+045 & 230.34 & 4.34 & 0.052 & I & 0.41 & 2.65 & 0.18 & 1.09\\
4FGL\,J1543.6+0452 & TXS\,1541+050 & 235.89 & 4.87 & 0.04 & I & 0.01 & 0.037 & 0.34 & 3.37\\
4FGL\,J1630.6+8234 & NGC\,6251 & 248.13 & 82.54 & 0.024 & I & 0.8 & 1.052 & 2.08 & 7.89\\
4FGL\,J1724.2$-$6501 & PKS\,1718$-$649 & 260.92 & $-$65.01 & 0.014 & ... & 0.57 & 0.266 & 0.57 & 1.68\\
4FGL\,J1843.4$-$4835 & PKS\,1839$-$48 & 280.81 & $-$48.61 & 0.111 & II & 5.19 & 163.895 & 0.14 & 0.98\\
4FGL\,J2156.0$-$6942 & PKS\,2153$-$69 & 329.27 & $-$69.69 & 0.028 & II & 30.5 & 55.864 & 0.79 & 2.21\\
4FGL\,J2227.9$-$3031 & PKS\,2225$-$308 & 336.98 & $-$30.58 & 0.058 & I & 0.1 & 0.823 & 0.13 & 0.92\\
4FGL\,J2302.8$-$1841 & PKS\,2300$-$18 & 345.76 & $-$18.69 & 0.129 & II & 0.86 & 37.783 & 0.3 & 1.43\\
4FGL\,J2326.9$-$0201 & PKS\,2324$-$02 & 351.72 & $-$2.04 & 0.188 & II & 0.14 & 14.32 & 0.65 & 2.01\\
4FGL\,J2329.7$-$2118 & PKS\,2327$-$215 & 352.42 & $-$21.3 & 0.031 & I & 0.65 & 1.411 & 0.94 & 2.91\\
4FGL\,J2341.8$-$2917 & TXS\,2338$-$295 & 355.37 & $-$29.32 & 0.052 & I & 0.04 & 0.266 & 0.28 & 1.18\\
\hline
\hline
\end{tabular}
\end{center}
\end{adjustwidth}
\end{table}

\section{Sample selection}
\label{sec:sample}
We have used the sample of radio galaxies presented in the 4th LAT AGN Catalog~\citep[4LAC,][]{4lac}, which includes a total of 41 sources. 
We have collected redshifts and 1.4\,GHz flux densities from the \href{http://ned.ipac.caltech.edu/}{NASA/IPAC Extragalactic Database (NED)}. Assuming a cosmology with $H_0 = 70$ km/s/Mpc and $\Omega_m=0.3$, we used the redshift to derive a luminosity distance and convert the radio flux density to power. We used these values to classify less-known sources as FR\,I or FR\,II, adopting a threshold of $L_\mathrm{1.4\,GHz} = 10^{25}$\,W/Hz~\citep{fan}. The final census includes 28 FR\,I and 12 FR\,II, in addition to the young radio galaxy PKS\,1718$-$649~\citep[see e.g.,][]{Migliori2016}. The prevalence of FR\,I radio galaxies among the $\gamma$-ray detected population is a well-known fact~\citep[see][]{magn}, which is yet to be explained.

\section {CTA simulated observations}
\label{sec:sim}
We have simulated CTA observations for all the radio galaxies in our sample, in order to evaluate their detection prospects, using the software package \texttt{ctools}~\citep{ctools}~\footnote{\href{http://cta.irap.omp.eu/ctools/}{http://cta.irap.omp.eu/ctools/}}. We extrapolated the \textit{Fermi}-LAT spectrum of the sources in the energy range 0.03-150 TeV and simulated 50 hours of CTA observations, which correspond to a standard single source observing campaign with Cherenkov arrays, using the \texttt{ctobssim} tool. We have used the coordinates of the two CTA sites to calculate the maximum altitude of each source from each site, and adopted the IRF corresponding to the site providing the maximum altitude. We then performed a standard likelihood analysis using the \texttt{ctlike} tool to derive a detection significance, in the form of a Test Statistic, which is defined as TS=$2\log(L/L_0)$ where $L$ is the likelihood of the model with a point source at the target position, and $L_0$ is the likelihood without the source. A value of TS=25 corresponds to a significance of 4.2$\sigma$~\citep{ts}.

We assumed as starting model the \textit{Fermi}-LAT spectrum in the 0.1-300 GeV range, as reported in the 4FGL~\citep{4FGL}. Out of all the radio galaxies in our sample, only NGC\,1275, 3C\,120, NGC\,6251 and PKS\,2153$-$69 show significant spectral curvature according to the 4FGL, where their spectrum is described with a LogParabola. All the other sources have PowerLaw spectra. 
We have chosen to adopt a PowerLaw spectrum as a base for the CTA simulations for 3C\,120 and PKS\,2153$-$69 as well, since these are faint sources where the curvature is not as well determined as for NGC\,1275 and NGC\,6251.Given that it is quite unrealistic to expect that a spectrum without curvature would extend from the LAT energy range down to 150 TeV (in the case of LAT sources with a PowerLaw spectrum), we have simulated different scenarios to account for spectral curvature, following \cite{2017APh....92...42A}. We assumed a PowerLaw with exponential cutoff at 0.1, 0.5 and 1 TeV, for all sources except NGC\,1275 and NGC\,6251, where a simple extrapolation of the \textit{Fermi}-LAT LogParabola model was adopted, since it already accounts for curvature. In all cases, we have included in the model only the target as a point source (see Section~\ref{sec:fornax} for additional discussions on the extension of Fornax\,A) and the CTA background.

We have considered the effect of $\gamma$-ray attenuation due to the Extragalactic Background Light~\citep[EBL, see e.g.,][]{EBL} for sources with $z>0.05$, i.e. 14 sources out of 41 (see Table~\ref{tab:sample}). To do so, we have added a multiplicative component to the spectral model, in the form $e^{-\tau}$. The values of the opacity $\tau$ as a function of energy and redshift have been collected from \cite{dom11}, and interpolated to the specific redshift of each source using a spline fit. The data from \cite{dom11} span the energy range 0.03-30 TeV, and therefore do not cover the range 30-150 TeV which is included in our simulations. However, at these energies the intrinsic spectrum is already steeply falling due to the applied exponential cutoffs, therefore including the EBL attenuation factor would not provide a significant difference.

\section{Results and discussion}
\label{sec:res}

\subsection{General properties}
The results of our simulated CTA observations are summarized in Table~\ref{tab:cta_sim}. For each source, we list the FR class (I or II), the normalization of the PowerLaw spectrum at 1 GeV, the photon index, the CTA site used for the simulations, and the resulting TS values for the cases described above. In our previous work, we found that the most realistic spectral shape is the one with $E_\mathrm{cut}=0.5$\,TeV. Based on our updated results, we can confirm this, by looking at current TeV radio galaxies (highlighted in bold in Table~\ref{tab:cta_sim}): while they are all confidently detected ($>5\sigma$) in the case $E_\mathrm{cut}=0.5$\,TeV, only two of them would be seen if $E_\mathrm{cut}=0.1$\,TeV. Looking at so far undetected candidates, we predict that the CTA will be able to detect eleven new radio galaxies with a significance $>5\sigma$ in 50 hours, assuming $E_\mathrm{cut}=0.5$\,TeV~\footnote{TXS\,0149+710 and NGC\,6251 have been excluded from this total because their high declination implies a low maximum altitude of ~47.5$^\circ$ and ~36.2$^\circ$, respectively.}. Of these, nine are FR\,I and two are FR\,II. If these predictions will be confirmed by future observations, the sample of TeV radio galaxies would be dramatically expanded, from the current six sources up to a total of seventeen. Moreover, since the six currently detected TeV radio galaxies are all FR\,Is, we can expect the CTA to allow us to study the FR\,I--FR\,II dichotomy in the VHE range for the first time, albeit still with a small FR\,II sample.

\begin{figure*}
    \centering
    \begin{adjustwidth}{-1.5cm}{-3.5cm}

    \includegraphics[width = 0.6\textwidth]{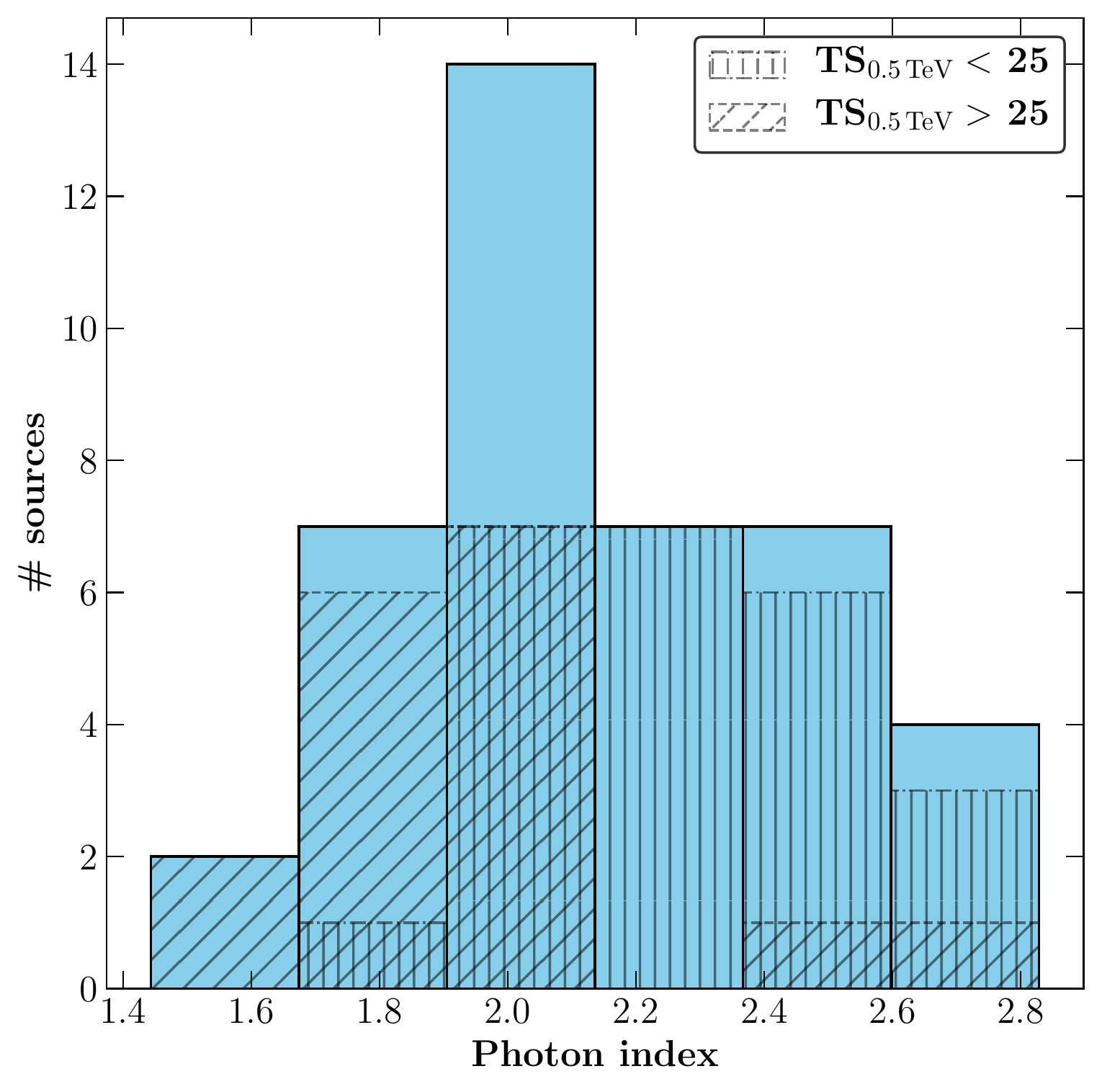}
    \includegraphics[width = 0.6\textwidth]{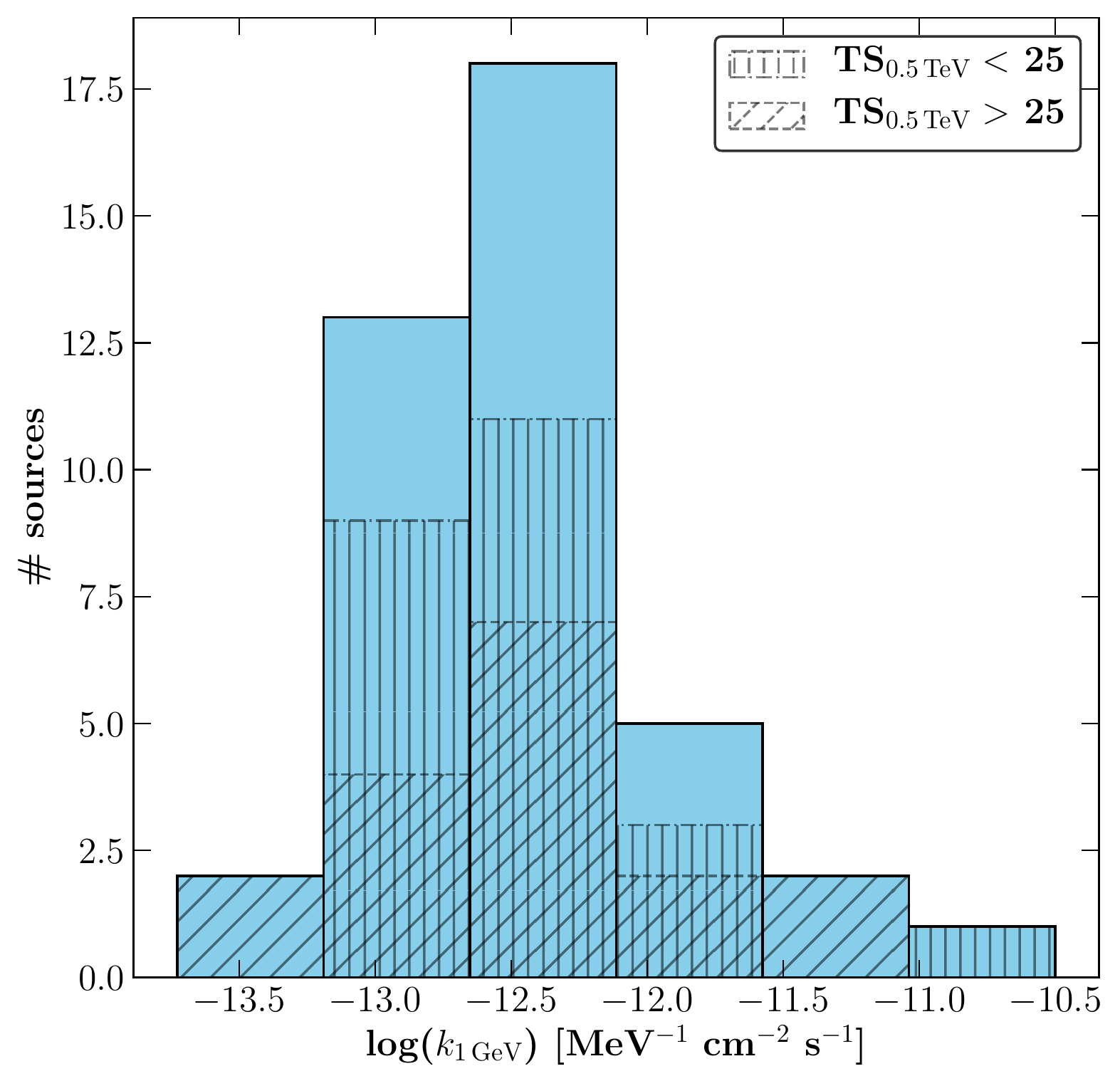}
    \end{adjustwidth} 
    \caption{Distribution of photon index and normalization for radio galaxies based on TS$_\mathrm{0.5 TeV}$. The light blue area indicates the total distribution while the hatched areas indicate the two subsamples with a threshold at TS$_\mathrm{0.5 TeV}=$~25.}
      
    \label{fig:hist}
\end{figure*}

We can see how the key parameter in determining the detectability of a source at TeV energies is the spectral slope, while the normalization plays a less significant role. We have quantified this by means of a two-sample Kolmogorov-Smirnov (KS) test, which shows that the distributions of sources with TS$_\mathrm{0.5 TeV}>$~25 and TS$_\mathrm{0.5 TeV}<$~25 differ significantly in photon index (KS=0.57, $p$-value=$1.5\times10^{-3}$), but not in normalization (KS=0.20, $p$-value=0.76). This can be seen in the histograms in Fig.~\ref{fig:hist}, where the distribution of photon index (left panel) is mostly bimodal, while the distribution of normalization (right panel) shows a much more significant overlap between the two subsets. These results confirm our previous findings reported in \cite{2017APh....92...42A}. We note that this does not mean that the spectral normalization is not a relevant selection criterion for a good TeV candidate, since very nearby sources will inevitably be easier to detect regardless of their spectral shape.

To further explore the best candidates, we have performed additional simulations of CTA observations with an integration time of five hours. Only potentially new TeV sources with a TS$\gtrsim$100 in the 50 hours observations with $E_\mathrm{cut}=0.5$\,TeV have been considered, i.e. 3C\,78, 4C\,+39.12, B2\,1113+29, Cen\,B, B2\,1447+27 and TXS\,1541+050. Not surprisingly, all of these sources are classified as FR\,I. Our likelihood analysis of the simulated five hours observations shows that 3C\,78 and TXS\,1541+050 can be detected with TS$>$25 even with short integration times, in the case $E_\mathrm{cut}=0.5$ TeV. This suggests that a VHE detection of these sources is most likely already within reach of current ground-based $\gamma$-ray observatories. 

\begin{figure}[htbp]
    \centering
    \includegraphics[width=0.8\linewidth]{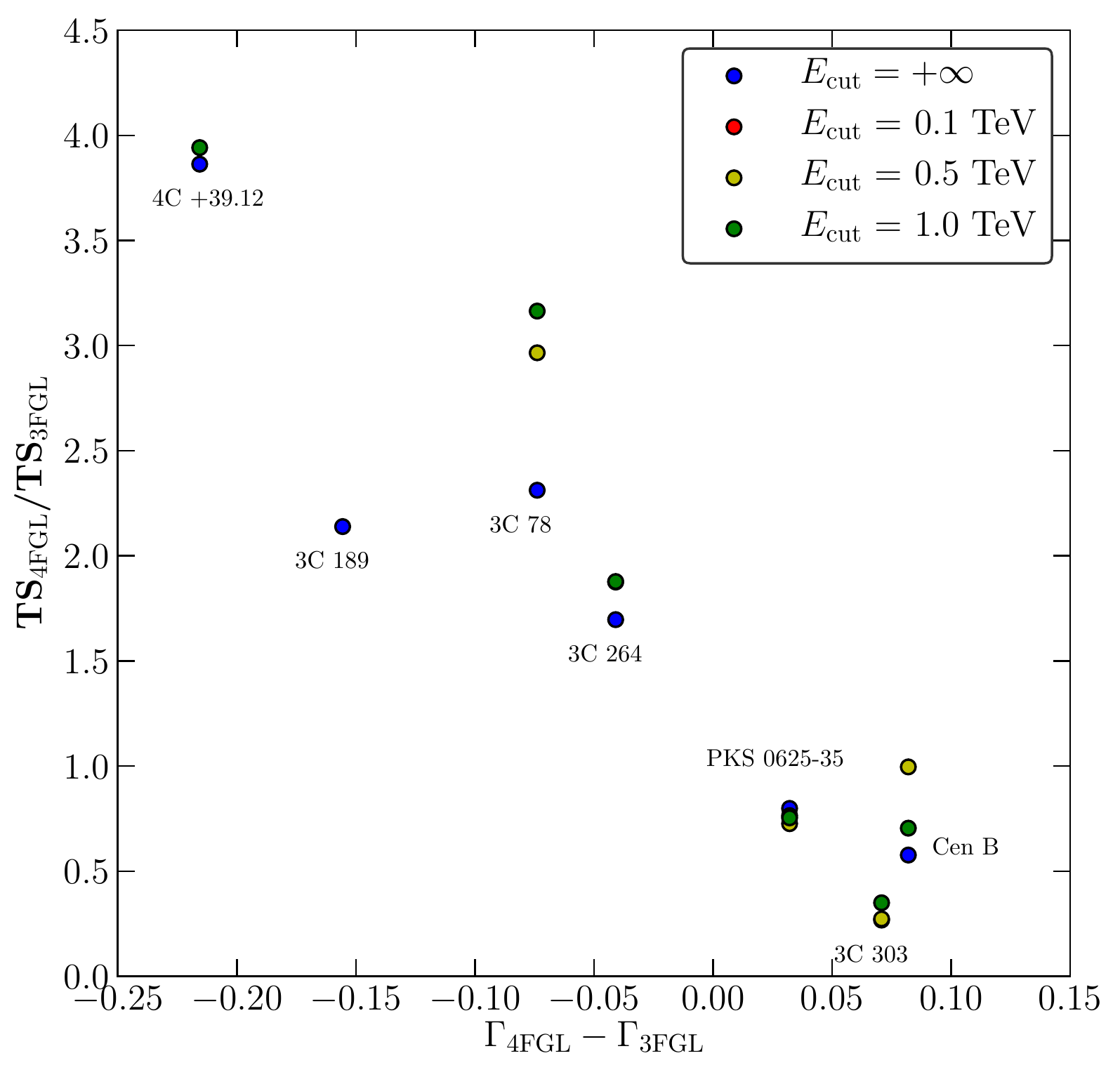}
    \caption{Scatter plot of ratio of simulated TS values with the CTA versus difference in PowerLaw photon index between the 3FGL and 4FGL spectral models. Negative values of $\Delta\Gamma=\Gamma_\mathrm{4FGL}-\Gamma_\mathrm{3FGL}$ indicate a harder spectrum in the 4FGL with respect to the 3FGL. Values of TS$_\mathrm{3FGL}$ are from Table\,3 of \cite{2017APh....92...42A}.}
    \label{fig:comparison}
\end{figure}

We note that, in several cases, our updated simulations yield significantly different results for sources included in our previous study~\citep{2017APh....92...42A}. This is most likely due to the change in the spectral parameters adopted for the extrapolation between the 3FGL and 4FGL catalogs. We tested this by computing Kendall's $\tau$ linear correlation test between the difference in photon index and the TS ratio between the 4FGL-based and 3FGL-based simulations (see Fig.~\ref{fig:comparison}). The test yields a significant anti-correlation, with $\tau$=-0.74 and a $p$-value=2$\times10^{-5}$, i.e. the 4FGL/3FGL TS ratio is higher when the source spectrum is harder in the 4FGL with respect to the 3FGL, as expected~\footnote{Fornax\,A has been excluded from this comparison since it was fitted as a point source in the 3FGL and as extended source in the 4FGL, see Section~\ref{sec:fornax}.}.

\subsection{Extended $\gamma$-ray emission from Fornax~A}
\label{sec:fornax}
The case of Fornax\,A deserves further investigation. In Table~\ref{tab:cta_sim} we have reported the results of CTA simulated observations assuming a point source spatial model for all sources. However, Fornax\,A has been established as the second extragalactic source with significantly detected extended $\gamma$-ray emission~\citep{2016ApJ...826....1A} after Centaurus\,A. Moreover, the core contribution to the GeV emission from Fornax\,A revealed by \textit{Fermi}-LAT was constrained to be $<14\%$. Therefore, a detection of this source in the E$>$100 GeV range would provide unique insight into the production mechanisms of the highest energy photons in radio galaxy lobes. 

The sensitivity of Cherenkov telescopes to extended emission is lower than for point sources. Therefore, we have performed additional simulated CTA observations of Fornax\,A, using two of the spatial models discussed in \cite{2016ApJ...826....1A} and an extrapolation of the 4FGL spectrum. Naturally, this implies a simplistic assumption that the spectral properties of the $\gamma$-ray emission are uniform across the source extension, which may well not be the case, and that the extension in the 0.03-150 TeV energy range is the same as in the 0.1-300 GeV range. The simplest extended model is a radial disk slightly offset from the catalog position, with a best-fit radius $r=0.33^\circ\pm0.05^\circ$~\citep{2016ApJ...826....1A}. The alternative spatial model is derived from the VLA 1.5\,GHz radio map from \cite{1989ApJ...346L..17F}, and is included in the 4FGL catalog~\footnote{See \href{https://fermi.gsfc.nasa.gov/ssc/data/access/lat/8yr_catalog/LAT_extended_sources_8years.tgz}{https://fermi.gsfc.nasa.gov/ssc/data/access/lat/8yr\_catalog/LAT\_extended\_sources\_8years.tgz}}.

We present the results of these additional tests in Table~\ref{tab:fornax} for integration times of 50 hours and 5 hours, together with the ones from the point source fit. For 50 hours of integration and $E_\mathrm{cut}=0.5$\,TeV, the significance of the radial disk and spatial map models are $\sim\!5$ times and $\sim\!15$ times smaller than the one for the point source model, respectively, but the radial disk model fit still yields TS$>$25, and the spatial map model is not far from this threshold. The spatial map model yields TS$>$25 in the case of $E_\mathrm{cut}=1.0$\,TeV, however this cannot be considered a likely scenario, since the \textit{Fermi}-LAT observed a highest photon energy of $\sim\!86$\,GeV from Fornax\,A, according to the third catalog of hard-spectrum sources \citep[3FHL, ][]{3fhl}. Finally, the simulations with 5 hours of integration time show that the CTA sensitivity to extended emission is not sufficient to significantly detect Fornax\,A on such short time scales, contrary to the results assuming a point source spatial model.

We can conclude from our simulations that a significant detection of extended $\gamma$-ray emission from Fornax\,A with the CTA appears possible, if a single-source campaign of $\gtrsim\!50$ hours integration time will be performed, under realistic assumptions on the source's spectral shape.


\begin{table*}
\caption{Results of simulated CTA observations of the radio galaxies in the sample, with 50 hours of integration. New $\gamma$-ray sources or reclassified sources (with respect to the 3FGL) are highlighted in italic. 
Currently known TeV radio galaxies are highlighted in bold.}
\label{tab:cta_sim}
\footnotesize
\tabcolsep=1mm
\begin{adjustwidth}{-3.5cm}{-3.5cm}

\begin{center}

\begin{tabular}{lccccccc}
\hline
\hline
 Name & FR class & $k^a$ & $\Gamma^b$ & CTA site & TS$_\mathrm{1 TeV}$ & TS$_\mathrm{0.5 TeV}$ & TS$_\mathrm{0.1 TeV}$ \\
\hline
\textit{IC 1531} &I & 22.10 & 2.20 & S & 28.1 & 14.7 & -0.0\\
\textit{NGC\, 315} &I & 41.95 & 2.34 & N & 0.0 & -0.0 & -0.0\\
\textit{TXS\,0149+710} &I & 30.41 & 1.90 & N & 557.4 & 195.8 & -2.0\\
\textit{PKS\,0235+017} &I & 15.97 & 2.17 & N & 0.0 & -0.0 & -0.0\\
3C\,78 &I & 70.19 & 2.00 & N & 810.0 & 284.7 & 8.0\\
\textit{NRAO\,128} &II & 42.66 & 2.47 & N & 0.5 & 0.6 & -0.0\\
\textbf{IC 310} &I & 18.29 & 1.78 & N & 819.6 & 269.7 & -1.4\\
\textbf{NGC\,1275$^c$} &I & 3161.23 & 2.12 & N & 127900616.7 & ... & ...\\
Fornax A &I & 58.50 & 2.05 & S & 780.0 & 346.0 & 15.2\\
4C\,+39.12 &I & 20.38 & 1.90 & N & 256.2 & 97.1 & -2.4\\
3C\,111 &II & 139.14 & 2.71 & N & -0.0 & -0.0 & -0.0\\
\textit{3C\,120$^d$} &I & 126.99 & 2.72 & N & -0.0 & -0.0 & -0.0\\
Pictor A &II & 48.39 & 2.46 & S & -0.0 & -0.0 & -0.0\\
\textbf{PKS\,0625$-$35} &I & 127.56 & 1.90 & S & 8818.4 & 4147.4 & 177.3\\
\textit{NGC\,2329} &I & 10.70 & 1.95 & N & 34.8 & 17.2 & -0.0\\
3C\,189 &II & 12.87 & 2.00 & N & 39.8 & 12.5 & -0.0\\
\textit{NGC\,2892} &I & 58.99 & 2.23 & N & 44.9 & 20.3 & -0.0\\
\textit{B3\,1009+427} &II & 23.05 & 1.76 & N & 42.6 & 28.0 & 1.4\\
\textit{B2\,1113+29} &I & 1.88 & 1.44 & N & 794.4 & 197.0 & -2.3\\
\textbf{3C\,264} &I & 27.07 & 1.94 & N & 240.3 & 82.5 & -2.2\\
\textit{NGC\,3894} &I & 23.24 & 2.06 & N & 66.9 & 16.2 & 1.9\\
\textbf{M 87} &I & 159.52 & 2.05 & N & 1589.6 & 702.0 & 29.3\\
\textit{TXS\,1303+114} &I & 14.35 & 1.95 & N & 36.4 & 21.1 & 0.5\\
\textit{PKS\,1304$-$215} &II & 57.20 & 2.13 & S & 96.9 & 58.9 & 2.3\\
\textbf{Centaurus A core} &I & 586.94 & 2.65 & S & 97.1 & 51.2 & -0.0\\
3C\,303 &II & 13.10 & 1.98 & N & 46.8 & 15.0 & -0.0\\
Centaurus B &I & 322.8 & 2.40 & S & 366.3 & 211.1 & 17.6\\
3C\,303 &II & 13.10 & 1.98 & N & 16.0 & 2.7 & 0.1\\
\textit{B2 1447+27} &I & 3.30 & 1.54 & N & 664.8 & 155.4 & -1.9\\
3C\,306 &I & 64.34 & 2.44 & N & -0.0 & 2.4 & -0.0\\
\textit{PKS\,1514+00} &I & 43.19 & 2.59 & S & 1.7 & 2.3 & -0.0\\
\textit{TXS\,1516+064} &II & 12.07 & 1.86 & N & 58.9 & 19.8 & 0.4\\
\textit{PKS\,B1518+045} &I & 16.61 & 2.05 & N & 20.0 & 8.9 & 0.1\\
\textit{TXS\,1541+050} &I & 40.55 & 1.87 & N & 1342.7 & 397.8 & 8.6\\
NGC\,6251$^c$ &I & 156.28 & 2.35 & N & 219703120.8 & ... & ...\\
\textit{PKS\,1718$-$649} & ... & 27.63 & 2.49 & S & -0.0 & -0.0 & -0.0\\
\textit{PKS\,1839$-$48} &II & 13.91 & 1.99 & S & 42.5 & 27.2 & 0.7\\
\textit{PKS\,2153$-$69$^d$} &II & 28.04 & 2.83 & S & -7.3 & -0.0 & -0.0\\
\textit{PKS\,2225$-$308} &I & 13.24 & 1.99 & S & 70.2 & 35.7 & 1.2\\
\textit{PKS\,2300$-$18} &II & 23.79 & 2.17 & S & 20.9 & 11.1 & 0.2\\
\textit{PKS\,2324$-$02} &II & 33.79 & 2.44 & S & -0.0 & 2.4 & 2.1\\
\textit{PKS\,2327$-$215} &I & 48.68 & 2.45 & S & -0.1 & -0.0 & -0.0\\
\textit{TXS\,2338$-$295} &I & 20.08 & 2.24 & S & 10.7 & 10.1 & 0.0\\
\hline
\hline
\end{tabular}
\end{center}
\end{adjustwidth}
$^a$ Normalization of the \textit{Fermi}-LAT PowerLaw spectrum at 1 GeV in units of $10^{-14}\,\mathrm{MeV}^{-1}\,\mathrm{cm}^{-2}\,\mathrm{s}^{-1}$.\\
$^b$ \textit{Fermi}-LAT PowerLaw photon index.\\
$^c$ Sources with a LogParabola \textit{Fermi}-LAT spectrum. No simulations including a cutoff have been performed for these objects since they already show spectral curvature.\\
$^d$ Sources with a LogParabola spectrum in the 4FGL, for which we have chosen to assume a PowerLaw spectrum as base for the simulations (see main text).

\end{table*}

\begin{table*}
\caption{Results of simulated CTA observations of the radio galaxies in the sample, with 5 hours of integration. Only potentially new TeV sources with TS$\gtrsim$100 in the 50 hours simulations with $E_\mathrm{cut}=0.5$ TeV have been considered (Fornax\,A is treated separately, see Table~\ref{tab:fornax}). New $\gamma$-ray sources or reclassified sources (with respect to the 3FGL) are highlighted in italic.}
\label{tab:cta_sim_5h}
\begin{adjustwidth}{-3.5cm}{-3.5cm}
\begin{center}
\begin{tabular}{lccccccc}
\hline
\hline
 Name & FR class & $k^a$ & $\Gamma^b$ & CTA site & TS$_\mathrm{1 TeV}$ & TS$_\mathrm{0.5 TeV}$ & TS$_\mathrm{0.1 TeV}$ \\
\hline
3C\,78 &I & 70.19 & 2.00 & N & 69.7 & 26.8 & 0.0\\
4C\,+39.12 &I & 20.38 & 1.90 & N & 30.7 & 9.6 & -5.1\\
\textit{B2\,1113+29} &I & 1.88 & 1.44 & N & 71.5 & 10.9 & 1.3\\
Centaurus B &I & 322.8 & 2.40 & S & 28.3 & 18.6 & -0.0\\
\textit{B2 1447+27} &I & 3.30 & 1.54 & N & 105.6 & 11.9 & 2.0\\
\textit{TXS\,1541+050} &I & 40.55 & 1.87 & N & 95.5 & 40.2 & 3.5\\
\hline
\hline
\end{tabular}
\end{center}
\end{adjustwidth}
$^a$ Normalization of the \textit{Fermi}-LAT PowerLaw spectrum at 1 GeV in units of $10^{-14}\,\mathrm{MeV}$\,$^{-1}\,\mathrm{cm}$\,$^{-2}\,\mathrm{s}$\,$^{-1}$.\\
$^b$ \textit{Fermi}-LAT PowerLaw photon index.\\

\end{table*}

\begin{table*}
\caption{Results of CTA simulated observations of Fornax\,A with different spatial models for 50 hours and 5 hours integration. The radial disk parameters are from ~\cite{2016ApJ...826....1A}. The spatial map template is from the 4FGL catalog~\citep{4FGL}.}
\label{tab:fornax}
\begin{center}
\begin{tabular}{clccc}
\hline
\hline
 Integration time & Model & TS$_\mathrm{1 TeV}$ & TS$_\mathrm{0.5 TeV}$ & TS$_\mathrm{0.1 TeV}$\\
\hline
 & Point source & 780.0 & 346.0 & 15.2\\
50\,h & Radial disk & 107.4 & 65.5 & 13.5\\
 & Spatial map & 80.3 & 22.4 & -0.0\\
\hline
 & Point source & 80.9 & 33.4 & -0.0\\
5\,h & Radial disk & 4.3 & 3.0 & 0.13\\
 & Spatial map & 9.5 & 1.1 & -0.0\\
\hline
\hline
\end{tabular}
\end{center}
\end{table*}

\section{Summary and conclusions}
\label{sec:conc}
In this work, we complement our previous findings on TeV detection prospects for radio galaxies with the upcoming Cherenkov Telescope Array (CTA). In \cite{2017APh....92...42A} we simulated CTA observations for local ($z<0.15$) radio galaxies in the Third \textit{Fermi}-LAT source catalog~\citep[3FGL,][]{3fgl}, including a total of 17 sources. In this paper, we report on improved simulations taking advantage of the updated radio galaxy sample from the Fourth \textit{Fermi}-LAT source catalog~\citep[4FGL,][]{4FGL}. The new catalog benefits from double the integration time, and includes 41 radio galaxies, out of which six have been currently detected at E$>$100\,GeV by ground-based $\gamma$-ray observatories.

We have simulated 50 hours of CTA observations for all the sources in this sample using the \texttt{ctools} package. We assumed different extrapolations of the \textit{Fermi}-LAT spectrum, and accounted for EBL attenuation for sources with $z>0.05$. In the case of a high-energy cutoff at $E_\mathrm{cut}=0.5$\,TeV, we predict that the CTA will be able to detect eleven new radio galaxies at a significance $>5\sigma$ in 50 hours. This number is roughly double the one reported in our previous work, which can be attributed to the commensurate increase in the candidate sample size. If the CTA were to indeed detect these objects, this would triple the number of TeV-detected radio galaxies. Moreover, we predict that two out of these eleven potentially new sources would be FR\,II objects, which would be the first to be detected in the TeV range, allowing us to study the bimodal radio galaxy population at the highest energies for the first time.

We have confirmed that the key parameter in determining a good TeV-candidate radio galaxy based on \textit{Fermi}-LAT data is the spectral slope, while the normalization of the spectrum does not play a key role. This strengthens our previous findings reported in \cite{2017APh....92...42A}, due to the much larger sample of candidate sources.

We have performed an additional set of simulation using only 5 hours of CTA integration time, in order to investigate the most promising TeV candidates in more depth. We have found that two FR\,I sources, i.e., 3C\,78 and TXS\,1541+050, would still be detected with $>5\sigma$ significance within 5 hours, under the assumption of $E_\mathrm{cut}=0.5$\,TeV. This implies that these two sources are very likely within reach of current-generation Cherenkov observatories, if observed for a full targeted campaign, i.e. with an integration time of a few tens of hours. Therefore, targeting these sources with current IACTs would be an ideal stepping stone, paving the way for the CTA era of TeV radio galaxy studies. 

We have investigated in detail the lobe-dominated FR\,I Fornax\,A, which has been revealed as an extended $\gamma$-ray source by \textit{Fermi}-LAT. Given that the core contribution to $\gamma$-rays detected by \textit{Fermi}-LAT has been constrained to be $<\!14\%$, a detection of Fornax\,A at TeV energies would provide unique insight into the acceleration and emission processes of particles of the highest energies in the diffuse lobes of radio galaxies. We have simulated CTA observations of Fornax\,A assuming a radial disk model and a spatial map model, following the results of \textit{Fermi}-LAT observations. Since the CTA sensitivity to extended emission is lower than for point sources, both these models yield a lower detection significance in all spectral models tested. Nonetheless, we have found that a CTA detection of extended emission from Fornax\,A is likely in the case of the radial disk model, and still plausible in the case of the spatial map model, for integration times $\gtrsim\!50$ hours. 


Another of our conclusions from \cite{2017APh....92...42A} is confirmed in this work: the exposure time per source envisioned in the current CTA extragalactic survey, of the order of a few hours, would not be sufficient to detect a large number of new TeV radio galaxies ($\sim2$). Such a goal would require long, targeted campaigns with exposure times of the order of several tens of hours.

Finally, we note that the number of predicted new TeV detections should be treated as a lower limit, since it does not account for several additional factors: variability, which might transform a poor candidate into a promising one during flares, especially if it exhibits a harder-when-brighter behavior, as has been observed in some radio galaxies~\citep[see e.g.,][]{Janiak2016}; additional emission components at high energies, such as the one detected in Centaurus\,A~\citep{cenA}; sources whose high-energy Spectral Energy Distribution (SED) peaks at TeV energies, which might be too faint to be detected by \textit{Fermi}-LAT.

In conclusion, we confirm our prediction that the CTA will significantly impact our understanding of TeV radio galaxies, allowing to probe the properties of the overall population thanks to a much larger sample. It will also allow, for the first time, to compare the VHE properties of FR\,I and FR\,II sources.

\section*{Acknowledgements}

RA thanks Paola Grandi and Eleonora Torresi for fruitful discussions, and Laura Vega Garc\'ia as the MPIfR internal referee. RA is also grateful to the journal referees for constructive reports that greatly improved the article. This research has made use of the CTA instrument response functions provided by the CTA Consortium and Observatory. This research made use of \texttt{ctools}, a community-developed analysis package for Imaging Air Cherenkov Telescope data. \texttt{ctools} is based on \texttt{GammaLib}, a community-developed toolbox for the high-level analysis of astronomical $\gamma$-ray data. This research made use of Astropy,\footnote{\href{http://www.astropy.org}{http://www.astropy.org}} a community-developed core Python package for Astronomy \citep{astropy:2013, astropy:2018}.

\section*{References}
\def\memsai{Memorie della Società Astronomica Italiana}
\def\aj{AJ}%
\def\actaa{Acta Astron.}%
\def\araa{ARA\&A}%
\def\apj{ApJ}%
\def\apjl{ApJ}%
\def\apjs{ApJS}%
\def\ao{Appl.~Opt.}%
\def\apss{Ap\&SS}%
\def\aap{A\&A}%
\def\aapr{A\&A~Rev.}%
\def\aaps{A\&AS}%
\def\azh{AZh}%
\def\baas{BAAS}%
\def\bac{Bull. astr. Inst. Czechosl.}%
\def\caa{Chinese Astron. Astrophys.}%
\def\cjaa{Chinese J. Astron. Astrophys.}%
\def\icarus{Icarus}%
\def\jcap{J. Cosmology Astropart. Phys.}%
\def\jrasc{JRASC}%
\def\mnras{MNRAS}%
\def\memras{MmRAS}%
\def\na{New A}%
\def\nar{New A Rev.}%
\def\pasa{PASA}%
\def\pra{Phys.~Rev.~A}%
\def\prb{Phys.~Rev.~B}%
\def\prc{Phys.~Rev.~C}%
\def\prd{Phys.~Rev.~D}%
\def\pre{Phys.~Rev.~E}%
\def\prl{Phys.~Rev.~Lett.}%
\def\pasp{PASP}%
\def\pasj{PASJ}%
\def\qjras{QJRAS}%
\def\rmxaa{Rev. Mexicana Astron. Astrofis.}%
\def\skytel{S\&T}%
\def\solphys{Sol.~Phys.}%
\def\sovast{Soviet~Ast.}%
\def\ssr{Space~Sci.~Rev.}%
\def\zap{ZAp}%
\def\nat{Nature}%
\def\iaucirc{IAU~Circ.}%
\def\aplett{Astrophys.~Lett.}%
\def\apspr{Astrophys.~Space~Phys.~Res.}%
\def\bain{Bull.~Astron.~Inst.~Netherlands}%
\def\fcp{Fund.~Cosmic~Phys.}%
\def\gca{Geochim.~Cosmochim.~Acta}%
\def\grl{Geophys.~Res.~Lett.}%
\def\jcp{J.~Chem.~Phys.}%
\def\jgr{J.~Geophys.~Res.}%
\def\jqsrt{J.~Quant.~Spec.~Radiat.~Transf.}%
\def\memsai{Mem.~Soc.~Astron.~Italiana}%
\def\nphysa{Nucl.~Phys.~A}%
\def\physrep{Phys.~Rep.}%
\def\physscr{Phys.~Scr}%
\def\planss{Planet.~Space~Sci.}%
\def\procspie{Proc.~SPIE}%
          
\bibliography{mybibfile}

\end{document}